\documentstyle[prl,aps,epsfig]{revtex}
\bibstyle{unsrt}
\tighten
\begin{document}

\twocolumn[\hsize\textwidth\columnwidth\hsize\csname
@twocolumnfalse\endcsname
\draft
\title{Coherent Control of Quantum Chaotic Diffusion}
\author{Jiangbin Gong and Paul Brumer}
\address{ Chemical Physics Theory Group,\\
University of Toronto\\ Toronto, Canada  M5S 3H6}
\date{May 30, 2000}
\maketitle

\begin{abstract}              
Extensive coherent control over quantum chaotic
diffusion using the kicked rotor model is demonstrated and its origin in
deviations from random matrix theory is identified.
Further, the extent
of control in the presence of external decoherence is established.
The results are relevant to both areas of quantum chaos and coherent control.
\end{abstract}

\pacs{PACS Numbers: 05.45.Gg, 05.45.Mt, 05.60.Gg, 32.80.Qk}
\vskip1pc]

The kicked rotor and its classical limit, the standard map, have long
served as paradigms for classical and quantum chaos\cite{casati}. The classical dynamics
shows characteristic diffusive energy growth whereas the quantum dynamics
shows similar chaotic short time behavior, followed by the suppression of diffusion
at longer times.
In this Letter, we demonstrate that the quantum features of the 
chaotic kicked rotor allow for
extensive coherent control\cite{brumer} over quantum chaotic diffusion, even
in the presence of modest decoherence. 
In particular, we show that 
quantum relaxation dynamics in the kicked rotor model is sensitive to the
coherence characteristics of the initial state, and that altering these 
characteristics  allows  for  control  over  the  energy diffusion. 
The extent of the controlled behavior  is vast, from strong suppression
to strong enhancement of diffusion.

Consider the kicked rotor whose Hamiltonian is given by 
\begin{equation}
H(\hat{L}, \theta, t)=\frac{\hat{L}^{2}}{2I}+\lambda \cos(\theta)\sum_{n}\delta(t/T-n),
\end{equation}
where $\hat{L}$ is the angular momentum operator, $\theta$ is 
the conjugate angle,  $I$ is the moment of inertia, $\lambda$
is the strength of the ``kicking field", and $T$ is the time interval between kicks.
The quantum time evolution operator $\hat{F}$ for times $(N-1/2)T$ to $(N+1/2)T$ is given by
\cite{casati}
\begin{equation}
\hat{F}=\exp[i\frac{\tau}{4}\frac{\partial^{2}}{\partial \theta^{2}}]\exp[-ik\cos(\theta)]
\exp[i\frac{\tau}{4}\frac{\partial^{2}}{\partial \theta^{2}}], 
\end{equation}
with dimensionless parameters
$\tau=\hbar T/I$ and $ k=\lambda T/\hbar$.
The classical limit \cite{casati} of this quantum map is given by the  standard map, which,
when expressed in terms of
dimensionless  variables $\theta$ and the scaled c-number angular momentum
$\tilde{L}=L\tau/\hbar$, takes the following form,
\begin{eqnarray}
\theta_{N}=\theta_{N-1}+ (\tilde{L}_{N}+\tilde{L}_{N-1})/2 \nonumber \\
\tilde{L}_{N}=\tilde{L}_{N-1}+ \kappa \sin(\theta_{N-1}+\tilde{L}_{N-1}/2),
\label{cmap}
\end{eqnarray}
where $\kappa=k\tau$, and $(\tilde{L}_{N}, \theta_{N})$ represents 
the phase space location of a classical
trajectory after $N$ kicks.  
The system is chaotic for $\kappa>\kappa_{cr}=0.9716...$. The 
resultant diffusion constant can be defined as 
the absorption rate of the average scaled rotational energy
$\tilde{E}\equiv \langle \tilde{L}^{2}\rangle /2$.
Comparing classical and quantum dynamics for typical initial classical
states  shows that quantum dynamics displays significant  
suppression of the  classical chaotic diffusion, i.e. 
the external field can only excite a finite number of unperturbed energy
levels \cite{shepelyansky}.

The fact that the rotor is a Hamiltonian system and the kick is coherent
implies that the system maintains its quantum phase throughout the evolution.
If this is the case then the system should be controllable via coherent
control\cite{brumer}, i.e. by using quantum interference phenomena to affect the
dynamics. To demonstrate this, and to examine the extent of possible control,
we consider the dynamics of states which are 
initially comprised of superpositions of
two arbitrary angular momentum eigenstates, $|m\rangle=
\exp(im\theta)/\sqrt{2\pi}$ and
$|n\rangle=\exp(in\theta)/\sqrt{2\pi}$.  Each of these eigenstates is classically 
allowed,
with a corresponding classical distribution function given by
$\rho_{m}^{c}(\theta, \tilde{L})=\delta_{\tilde{L}/\tau, m}/2\pi$ and
$\rho_{n}^{c}(\theta, \tilde{L})=\delta_{\tilde{L}/\tau, n}/2\pi$,
respectively\cite{jaffe}. 

To show that changing the coherent characteristics of the initial state 
significantly alters the subsequent dynamics, we 
consider the dynamics of states given initially by the superposition
$ |\psi\rangle=\cos(\alpha)|m\rangle+\sin(\alpha)\exp(i\beta)|n\rangle $.
Typical results, culled from numerous cases of varying 
$\alpha,\beta$, $k$ and $\tau$ are shown below and
correspond to a weaker and stronger chaotic case, 
and to two values of $\beta$, i.e. $\beta=0,$ and $\beta=\pi$.
Specifically, we display
below results for case (a) $|\psi_{a}^{\pm}\rangle=
(|+2\rangle\pm |-1\rangle)/\sqrt{2}$, with $\tau=0.5$, $k=5.0$ and
for case (b) $|\psi_{b}^{\pm}\rangle=
(|+1\rangle\pm |+2\rangle)/\sqrt{2}$, with $\tau=1.0$, $k=5.0$.
Note that neither the basis states nor the superposition
states are eigenstates of the parity operator.

Figure 1 shows $\tilde{E}$ for each of these two systems
and for each of the values of $\beta$.  Figure 1a,
for example, displays $\tilde{E}$ for 
$|\psi_{a}^{-}\rangle$ ( dashed curve) and for
$|\psi_{a}^{+}\rangle$ (solid curve). Clearly, the initial state
$|\psi_{a}^{-}\rangle$  gives clear diffusive behavior 
during the first 40 kicks whereas energy absorption in the case of 
$|\psi_{a}^{+}\rangle$ is completely suppressed. 
As a result, $\tilde{E}(t=40T)=9.6$ for
the $|\psi_{a}^{-}\rangle$ case, while $\tilde{E}(t=40T)= 1.6$ 
for propagation from the 
initial state $|\psi_{a}^{+}\rangle$.  Note  (a) that this huge difference
is achieved solely by changing the initial relative phase $\beta$ between 
the two participating states $|+2\rangle$ and $|-1\rangle$
in the initial superposition, and (b) that by contrast, each of 
$|+2\rangle$ or $|-1\rangle$ behave similarly to one another with respect
to energy absorption, giving $\tilde{E}(t=40T)=5.4$ and 6.0, respectively.
Hence, the observed control is due entirely to changing the coherent 
properties of the initial superposition state.

\begin{figure}
\centerline{\epsfig{file=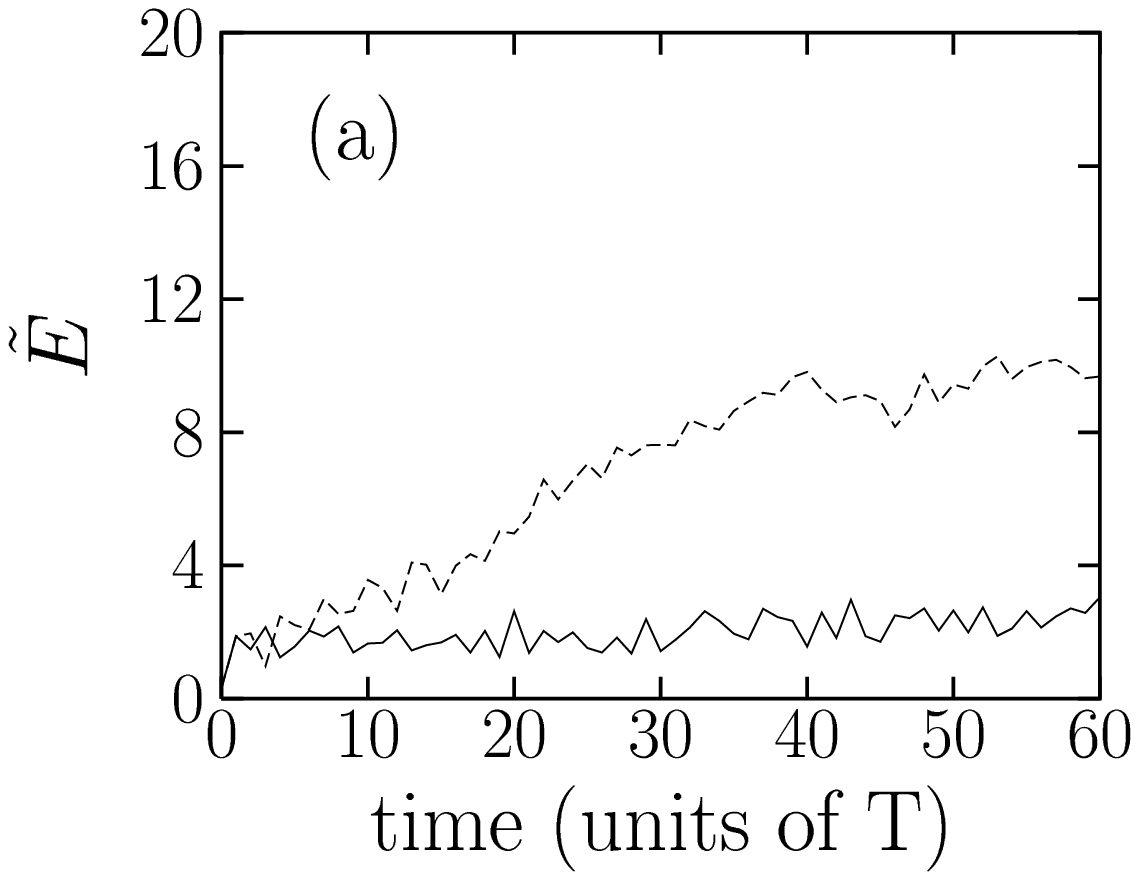,width=5.cm}}

\centerline{\epsfig{file=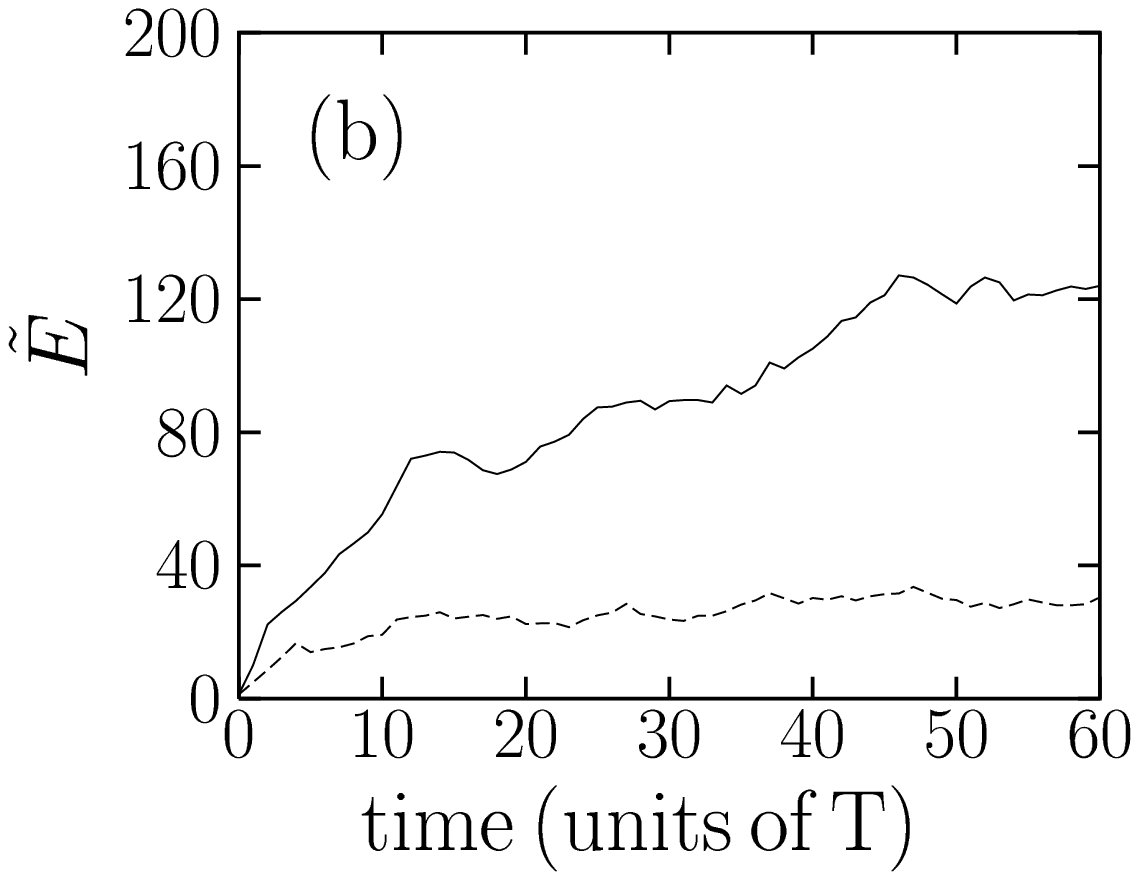,width=5.cm}}
\caption{
The expectation value of the dimensionless scaled rotational energy
$\tilde{E}=\langle\hat{L}^{2}\rangle \tau^{2}/2\hbar^{2}$ versus time (in units
of $T$). (a)
Solid and dashed lines are for the initial states $|\psi_{a}^{+}\rangle$
and $|\psi_{a}^{-}\rangle$, respectively,
$\tau=0.5$, $k=5.0$.
(b) Solid and dashed lines are for the initial states
$|\psi_{b}^{+}\rangle$ and $|\psi_{b}^{-}\rangle$, respectively,
$\tau=1.0$, $k=5.0$.}
\end{figure}

Similar control persists for the more chaotic case shown in Fig. 1b.
Here $|\psi_{b}^{+}\rangle$ (solid line) shows
extensive chaotic diffusion (i.e. compare ordinates scale for
Figs. 1a and 1b) for up to $45$ kicks, giving $\tilde{E}(t=45T)$ far higher than the
value of 70.4 and 77.1 reached by propagating either of the basis functions
$|+1\rangle$ and $|+2\rangle$. Further, and 
by contrast, there is essentially no quantum
diffusion after $t=4T$ for $|\psi_{b}^{-}\rangle$ (dashed line). 
Control (not shown) is possible for the resonant case as well, e.g., where
$\tau=\pi/3$, but it is somewhat less extensive.

These differences are also reflected in 
the details of the evolving wavefunctions. For example, Fig. 2 shows
the probability $P(m)$ of finding the system in the
state $|m\rangle$ at $t=60T$. For case (a), $P(m)$ for 
$|m|\geq 10$ is $15.8\%$ and $3.4\%$  for $\beta=\pi$ and for $\beta=0$,
respectively. Similarly, for case (b) 
$P(m)$ differs by a factor of $5.4$ for the two $\beta$ values
(3.2\% vs. 17.2\%)  in the
probability of exciting the rotor to high-energy
rotational states $|m\rangle$, $|m|\ge 20$. In both Figs. 2a and 2b it is
evident that the difference in final populations resulting from the evolution
of the two superpositions is an erratic function of $m$ with few evident
trends.

\begin{figure}

\vspace{0.8cm}
\centerline{\epsfig{file=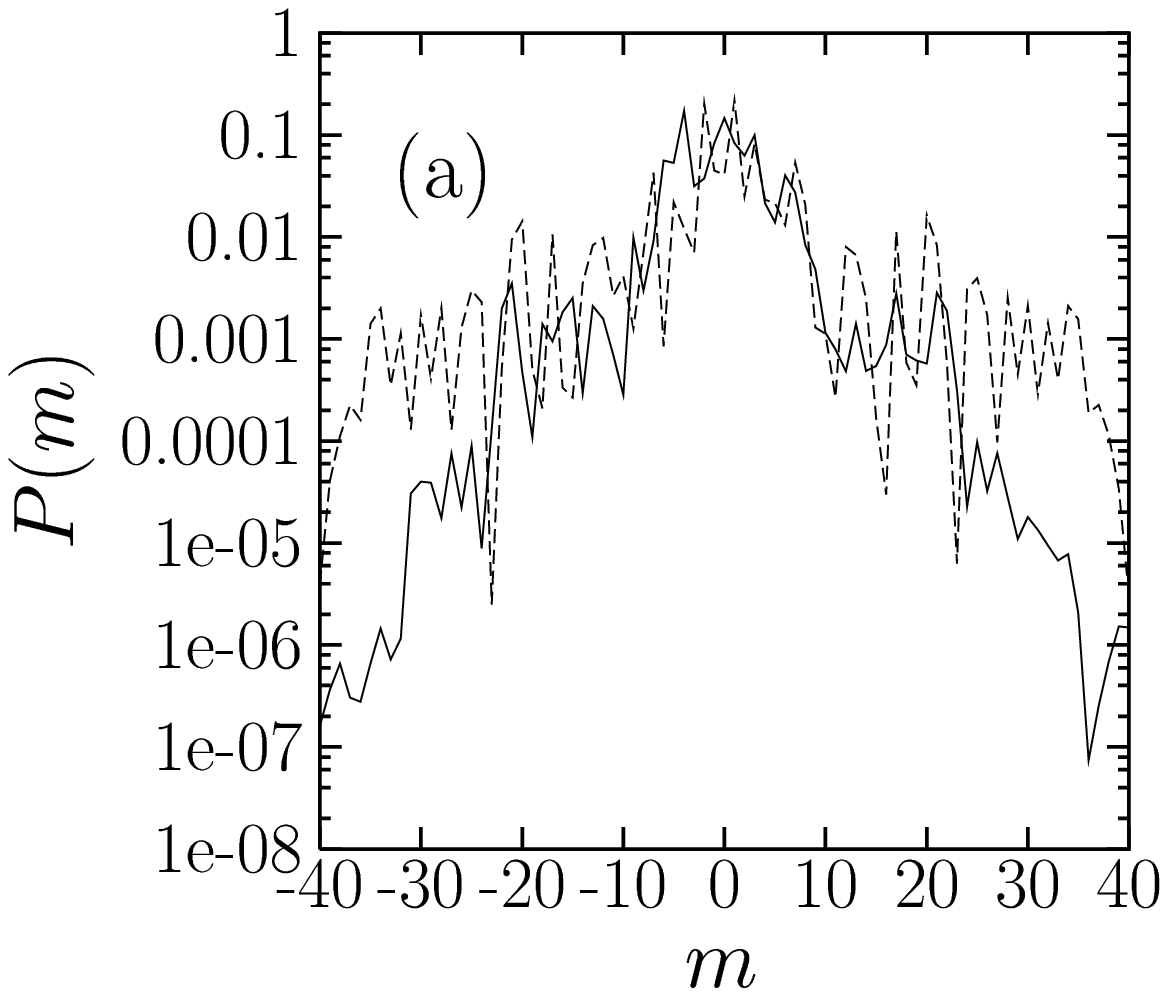,width=5.0cm}}

\vspace{0.6cm}
\centerline{\epsfig{file=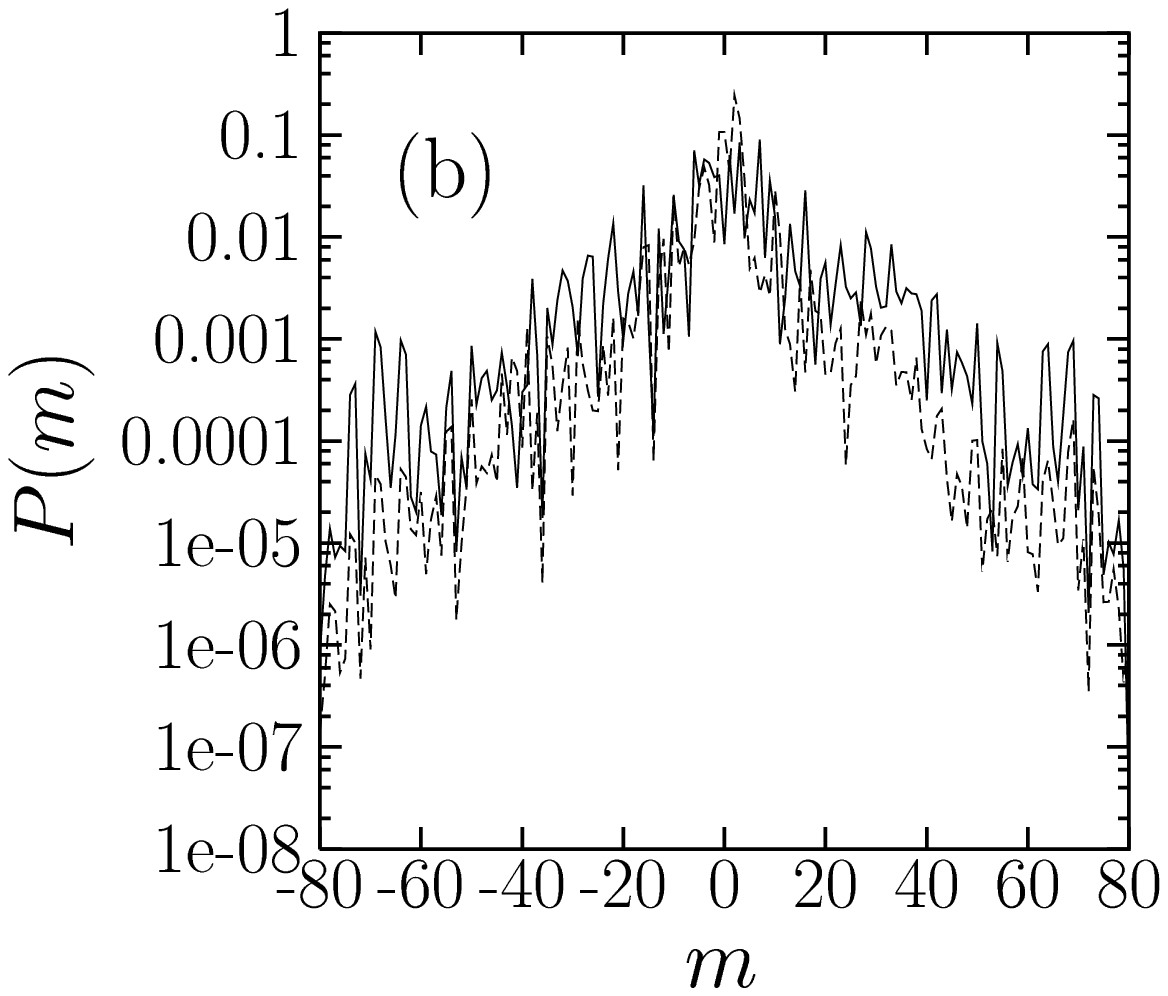,width=5.0cm}}
\caption{Probability $P(m)$ of finding the system in the state $|m\rangle$
  at $t=60T$. Results are for the cases shown in Figure 1}
  \end{figure}

The behavior shown in Fig. 1 is in sharp contrast to that which would be
observed for the same initial distributions propagated classically.
These computations are shown in Fig. 3 and result from classical propagation of
the initially  non-positive-definite  Wigner  function 
$\rho^{W}(\theta, \tilde{L})$    associated    with the
wavefunction $\cos(\alpha) |m\rangle + \sin(\alpha)\exp(i \beta) |n\rangle.$
That is, we classically propagate
\begin{eqnarray}
\rho^{W}(\theta, \tilde{L})
&=&\cos^{2}(\alpha)\rho_{m}^{c}(\theta, \tilde{L})
+\sin^{2}(\alpha)\rho_{n}^{c}(\theta, \tilde{L}) \nonumber \\
&&+\frac{1}{2\pi} \sin(2\alpha)\cos(\beta-(m-n)\theta)\delta_{\tilde{L}/\tau, (m+n)/2}.
\label{wigner}
\end{eqnarray}
for each of $|\psi_a^{\pm}\rangle$ and $|\psi_b^{\pm}\rangle$.
In all cases, the classical results show [Fig. 3] strong diffusion, characteristic of
the chaotic dynamics of the standard map. There are only small differences 
in the $\tilde{E}$ diffusion between $|\psi_a^{+}\rangle$ and
$|\psi_a^{-}\rangle$ and between $|\psi_b^{+}\rangle$ and $|\psi_b^{-}\rangle$.

Consider then the origins of coherent control of chaotic systems in the
quantum dynamics, and the behavior in the classical limit. To this end we
diagonalize the quantum map operator $\hat{F}$ by a unitary operator $\hat{U}$, i.e.,
$\langle i|\hat{F}|j\rangle= \sum_{k}e^{-i\phi_{k}}
U_{ki}^{*}U_{kj}$, where $ U_{ij}\equiv \langle i|\hat{U}|j\rangle, j=1,2,
\cdots $ is the eigenvector with eigenphase $\phi_{i}$.
After the initial superposition state $|\psi\rangle= \cos(\alpha) |m\rangle+
\sin(\alpha) \exp(i \beta) |n\rangle$  is kicked $N$ times,
we have
\begin{eqnarray}
\frac{2\tilde{E}}{\tau^{2}}&=&\cos^{2}(\alpha)\sum_{ljj'}l^{2}U_{jm}^{*}U_{j'l}^{*}
U_{jl}U_{j'm}e^{iN(\phi_{j}-\phi_{j'})} \nonumber \\
& &+ \sin^{2}(\alpha)\sum_{ljj'}l^{2}U_{jn}^{*}U_{j'l}^{*}
U_{jl}U_{j'n}e^{iN(\phi_{j}-\phi_{j'})} \nonumber \\
&&+[\frac{1}{2}\sin(2\alpha)e^{-i\beta}\sum_{ljj'}l^{2}
U_{jm}^{*}U_{j'n}U_{jl}U_{j'l}^{*}e^{iN(\phi_{j}-\phi_{j'})} 
\nonumber \\
&& + c.c.]
\label{prop}
\end{eqnarray}
where $c.c.$ denotes the complex conjugate of the immediately preceding term
within the brackets.
The total term in brackets corresponds to interference effects due to initial-state
coherence. For large $N$
only the $j=j'$ terms will survive in the summations due to rapid oscillations
of $e^{iN(\phi_{j}-\phi_{j'})}$. Hence 
the last two terms reduce to $1/2\sin(2\alpha)e^{-i\beta}
\sum_{l}l^{2} \sum_{j} |U_{jl}|^{2} U_{jm}^{*}U_{jn} + c.c.$.
\begin{figure}

\centerline{\epsfig{file=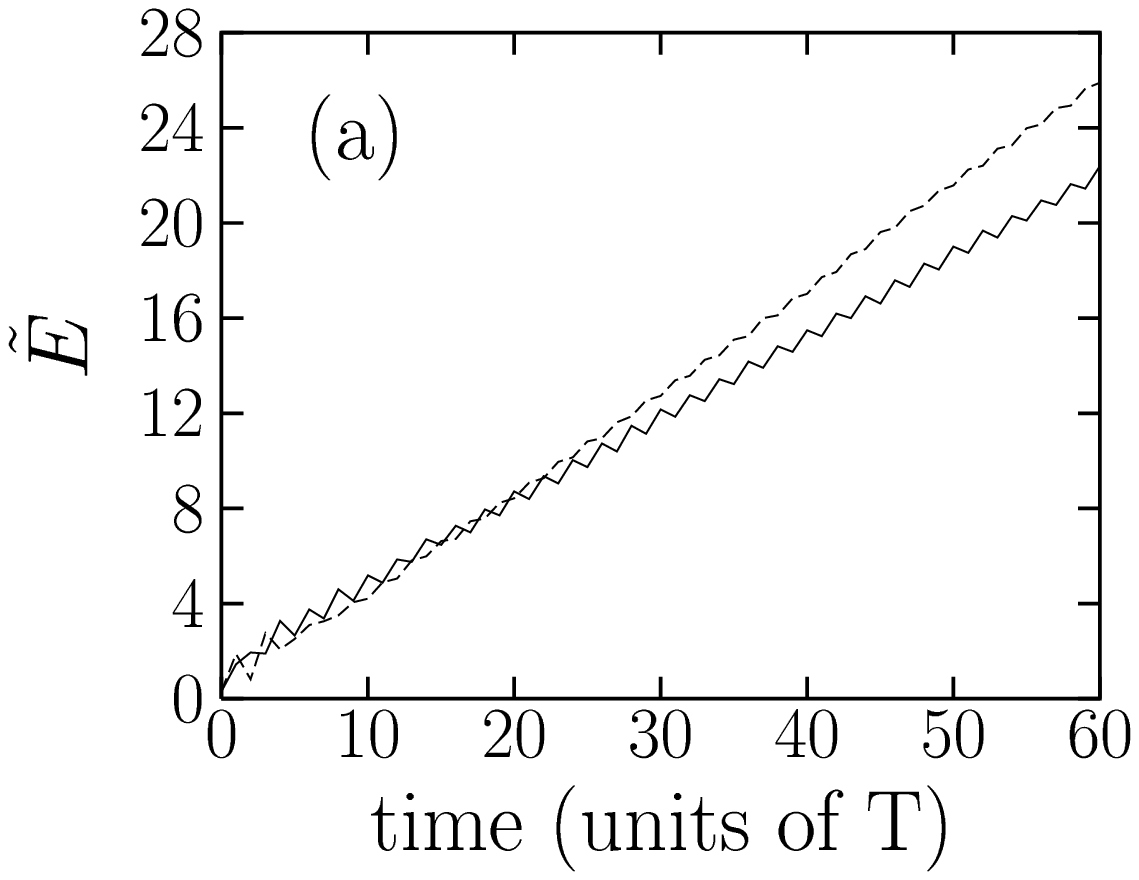,width=5.cm}}

\centerline{\epsfig{file=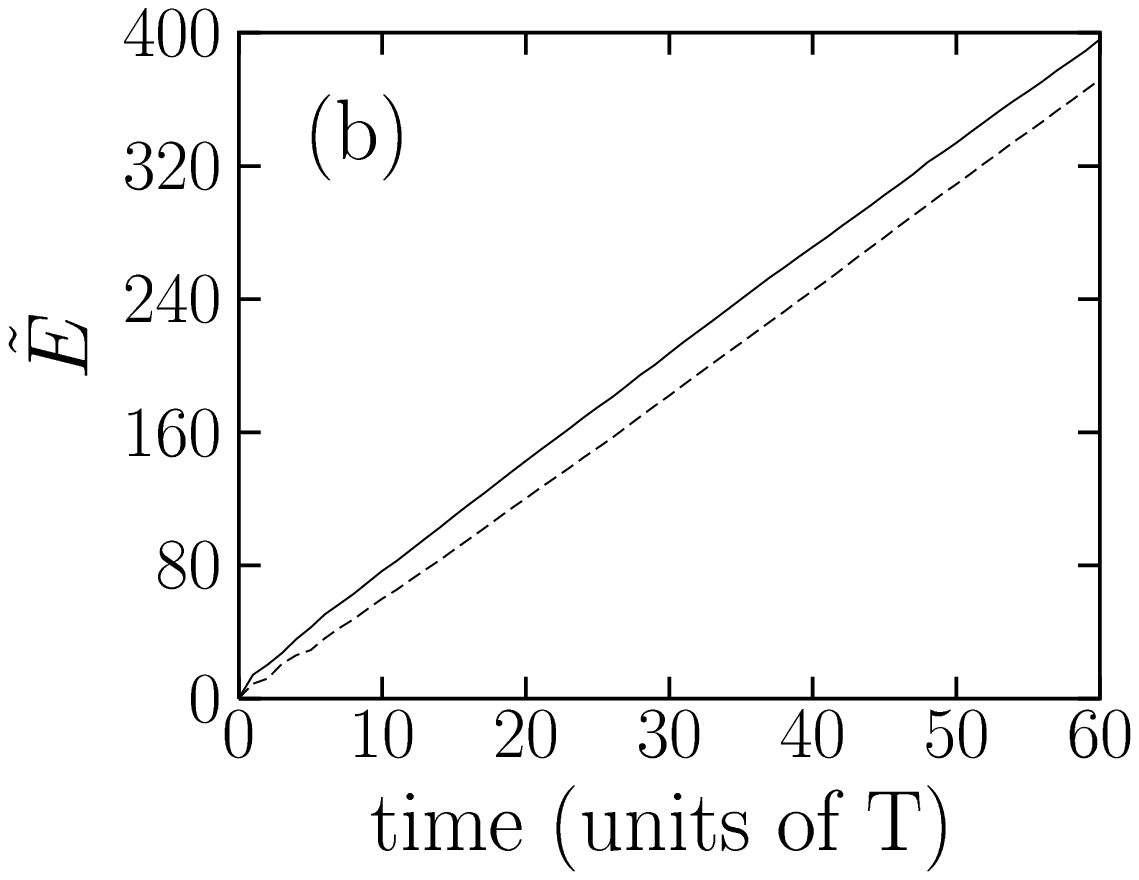,width=5.cm}}
\caption{As in Fig. 1 except that $\tilde{E}$\ is calculated by
  classically propagating the initial non-positive-definite Wigner function
	in Eq. (\ref{wigner}).}
	      \end{figure}

If there is no structure in $|U_{jl}|^{2}$ and
the eigenvector components  $U_{jm}^{*}$ and $U_{jn}$ are perfectly
independent, as expected from Random-Matrix-Theory\cite{haake}, then 
$\sum_{j} |U_{jl}|^{2} U_{jm}^{*}U_{jn}$ is small and
the magnitude of the interference term is $1/\sqrt{D}$ times smaller
than that of the incoherent terms (where $D$ is the effective dimension
of the Hilbert space)
\cite{csi},
interference vanishes and control is lost. 
Hence, the coherent  control of quantum chaotic diffusion relies upon 
the residual statistical correlations
between  eigenvector components  $\{U_{ij}\}$.  Indeed, in this system the 
matrix $\langle i|\hat{F}|j\rangle$ 
is known to 
display a band structure with the bandwidth $2k$ 
where the quantity
$k^2/N$, where $N$ is the size of the banded random matrices,
provides   a   measure  of  the  statistical  deviations  from  Random  Matrix
Theory\cite{izrailev}. Numerical results show that  
$k^{2}/N<0.2$ is sufficiently small for control to persist. This being the case,
we obtain a necessary condition to ensure the significance of the interference
term, namely, $k<0.2N_{\kappa}/\kappa $ or $ \tau> \kappa^{2}/(0.2N_{\kappa})$, where $N_{\kappa}$
is the minimum grid size for accurate FFT calculations with  $\tau=1$.
Numerical studies indicate that, for  $\kappa<10.0$, 
$N_{\kappa} \approx 256$, implying that we require 
$k<50/\kappa$ or $\tau>\kappa^{2}/50$ for control. This makes clear that as
one approaches the classical limit (by increasing $k$ or decreasing $\tau$ 
with fixed $\kappa$), coherent control is lost. 

Further evidence that deviations from random matrix theory are responsible 
for control was obtained by examining control using a model composed of a
banded matrix with random matrix elements. Control was obtained in this
case as well, but was not as extensive as the kicked rotor system since
the latter displays less random matrix character.

The dynamics of the kicked rotor in the presence of decoherence effects 
has also been examined both experimentally\cite{raizen} and theoretically
\cite{ott}.  The  survival  of  control in the presence of decoherence is of
interest both in general, and for this particular case. 
To examine this issue we introduce a simple decoherence model. 
Here,  the quantum map operator 
between $(N-1/2)T$ and $(N+1/2)T$ is taken as $\hat{R}\hat{F}$, where
$\hat{R}$ introduces  random phases into the system. Specifically,
$\hat{R}|m\rangle=e^{i2\pi r\xi(m,N)}|m\rangle$, where 
$m =0,\pm 1, \pm2, \cdots $,
and $\xi(m,N)$ takes on random values that are
distributed uniformly between $0$ and $1$ for
each different $m$ or $N$. Note that this model is such that 
its $r \rightarrow 1$ limit corresponds to
measurement-induced quantum diffusion\cite{facchi}.  

The density matrix $\hat{\rho}$ for the dynamics governed by $\hat{R}\hat{F}$
can then be obtained as an average over many realizations of $\xi(m,N)$. We take  
the linear entropy $S\equiv Tr\hat{\rho}^{2}$ as a useful additional
measure of the purity of quantum  states, and hence of the effect of
decoherence.

Numerical studies show that for  $r<0.05$,  coherent control of
quantum  diffusion is  hardly  affected by the decoherence.  For  stronger
decoherence, e.g. $r=0.15$, phase control is essentially lost.
Examination of the corresponding
values of $S(t=60T)$ shows that this is consistent with maintenance of
control when the decoherence is sufficiently small so that $S(t=60T)> 0.4$. 
Sample results are shown in Fig. 4
where  we plot the time dependence $\tilde{E}$ versus the time
in the presence of small decoherence [Fig. 4a, $r= 0.05$] and of
appreciable  decoherence  [Fig. 4b, $r=0.15$]. Both panels refer to the
case of stronger chaos [case (b)]. Comparison of
Fig. 4a with the decoherence-free dynamics in Fig. 1b shows that control 
is still significant, but decoherence is beginning to have an effect insofar as
$\tilde{E}$ at $t=60T$ is larger in Fig. 4a than in Fig. 1b. That is, 
the results show a slight
tendency towards the classical behavior. In the case of stronger
decoherence [Fig. 4b] phase control is greatly reduced and long time linear 
diffusive growth of $\tilde{E}$ is observed. A careful examination of
Fig. 4b suggests that phase control persists until $t \approx 20T$ when the
slopes of the dashed and solid curve become virtually identical. However,
the slopes of these curves are still significantly less than those in Fig. 3b,
suggesting that quantum coherence is still maintained at these longer times.
In essence, it appears that phase control over the diffusion rate
vanishes before quantum coherence
is completely destroyed. 

\begin{figure}

\centerline{\epsfig{file=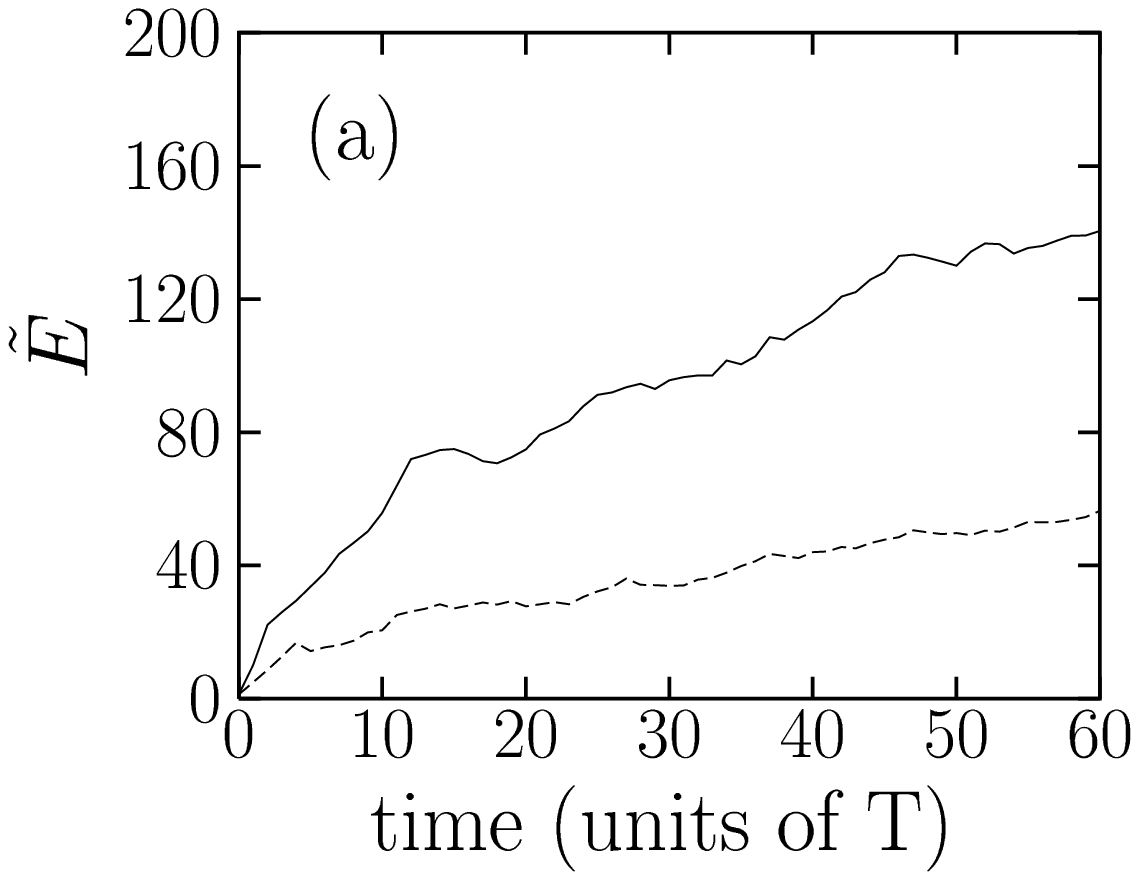,width=5.cm}}

\centerline{\epsfig{file=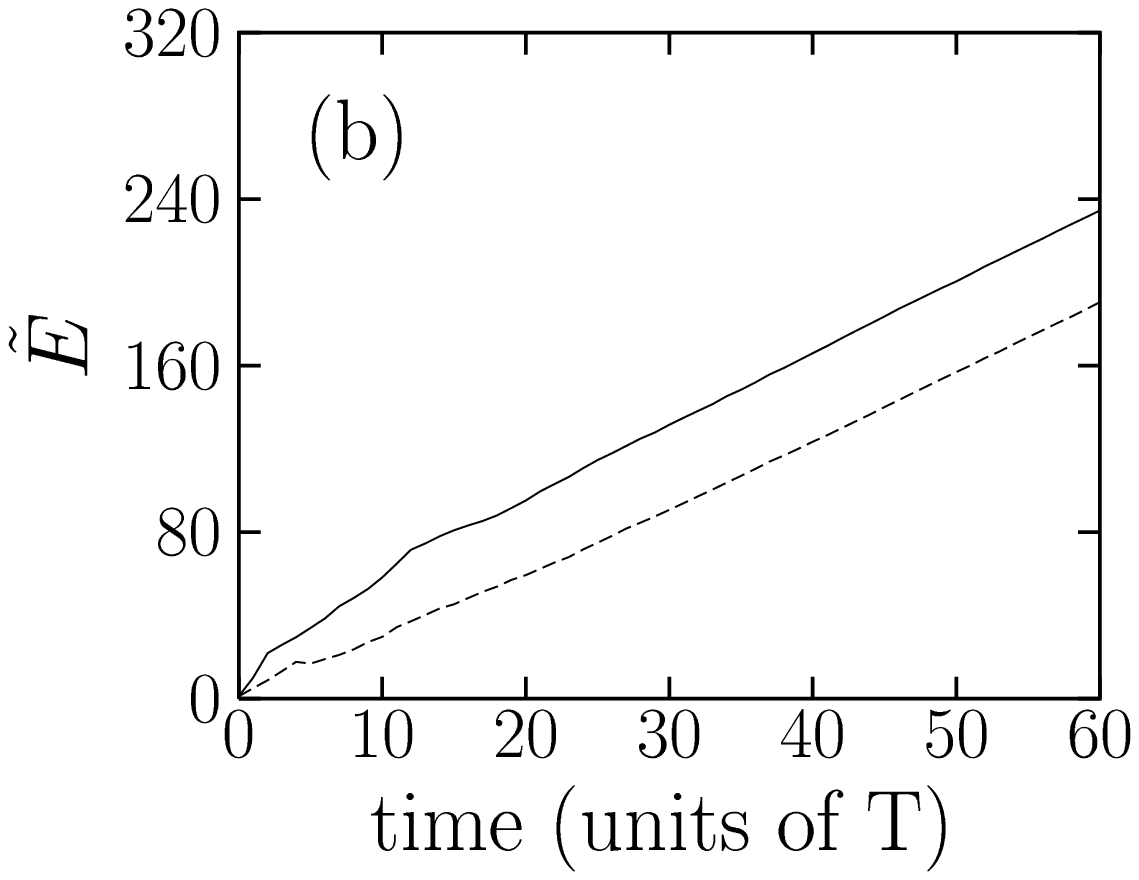,width=5.cm}}
\caption{As in Fig. 1b but in the presence of (a) modest decoherence and
  (b) stronger decoherence.}

	\end{figure}

Note, finally, that the possibility of  control 
does not rely heavily on the specific choice of
basis states \cite{parityfoot}. 
For example, we have also obtained extensive phase control
by adding together either negative parity real basis states 
$\sin(m\theta)/\sqrt{\pi}$ or by adding together positive parity states
$\cos(m\theta)/\sqrt{\pi}$.
Since the states $\sin(m\theta)/\sqrt{\pi}$
satisfy the boundary condition of an infinitely deep square-well potential $V(\theta)$ with
$V(0)=V(2\pi)=+\infty$ and since the dynamics of a kicked particle in a well
is similar to the kicked rotor for similar initial states\cite{well},
this indicates that one can also demonstrate 
coherent control of chaotic diffusion using superpositions of Hamiltonian
eigenstates of a kicked particle in a potential well.

A number of possible experimental demonstrations of 
the proposed control scenario are evident. For example, 
the kicked diatomic molecule $CsI$ \cite{csi,fishman} is a promising
molecular system for demonstrating controlled quantum chaotic diffusion.
In this case, preliminary controlled laser excitation could be used to prepare
the   desired   initial   superposition  state  (which are here superpositions
of $|J,M\rangle$ and $|J',M\rangle$, where $J$ and $J'$ are the angular momentum and
$M$ is their projection on the $z$ axis)
and to vary $\beta$. Alternatively, one can
utilize the square well analogy described above to experimentally study 
kicked dynamics of a particle in a well. By contrast, implementation of atom-optics
approach \cite{graham} to studies of control appears more difficult, insofar as it is necessary to 
prepare initial quantum superposition states, a considerable extension of
previous work \cite{raizen2}.

In summary, we have shown that quantum chaotic dynamics in systems such as the
kicked rotor may be controlled by varying, in accord with 
coherent control, the coherent characteristics of the initial state.
In particular, preparing even a simple superposition state comprising
two basis states allows a wide range of control over the diffusive dynamics.
The control persists in the presence of weak decoherence, but, as expected,
is lost with increasing decoherence strengths.
The predicted range of  control is extensive, providing results of broad
theoretical and experimental interest.

\vspace{0.2in}
{\bf Acknowledgments:} This work was supported by the U.S. Office of Naval Research
and the Natural Sciences and Engineering 
Research Council of Canada. We thank Professor Aephraim Steinberg for 
discussions on the atom-optics approach to $\delta$ kicked dynamics.

\pagebreak


\begin{references}
\bibitem{parityfoot} We have also obtained similar control results with $\cos \theta$
in Eq. (1) replaced by $\cos (\theta + \pi/10)$, which gives a Hamiltonian 
that does not conserve parity.
\end{references}
  \end{document}